\documentclass[prc,twocolumn,showpacs,showkeys,preprintnumbers,amsmath,amssymb,nofootinbib]{revtex4}
 \usepackage{graphicx}

\newcommand{\KNsing}{$\{\bar KN\}_{I=0}$}
\newcommand{\KK}[1]{$\{K\bar K\}_{I=#1}$}

\begin{document}
\title{$K\bar{K}N$ molecule
state with $I=1/2$ and $J^P=1/2^+$ studied with three-body calculation}

\author{Daisuke Jido and Yoshiko Kanada-En'yo}
\affiliation{Yukawa Institute for Theoretical Physics, Kyoto University,
Kyoto 606-8502, Japan}

\begin{abstract}
A $K\bar{K}N$ system with $I=1/2$ and $J^P=1/2^+$ 
is investigated with 
non-relativistic 
three-body calculations by using effective $\bar{K}N$, 
$K\bar K$ and $KN$ interactions. The 
$\bar{K}N$ interaction describes the 
$\Lambda(1405)$ as a $\bar{K}N$ molecule, and
the $K\bar K$ interaction is adjusted to give $f_0(980)$ and
$a_0(980)$ states as $K\bar K$ molecules.
The present investigation suggests that 
a bound $K\bar{K}N$ state can be formed below the 
$K\bar{K}N$ threshold (1930 MeV) 
with a $90 \sim 100$ MeV width of three-hadron decays, 
which are dominated by
$K\bar{K}N\rightarrow K\pi \Sigma$ and $\pi\eta N$. 
It is found that the $K\bar{K}N$ state is a weakly bound hadron molecular
state with a size larger than an $\alpha$ particle 
because of the repulsive $KN$ interactions.
\end{abstract}
\preprint{YITP-08-53}
\maketitle

\noindent

\section{Introduction}

Exploring composite systems of mesons and baryons is a challenging 
issue both in theoretical and experimental hadron-nuclear physics.
One of the historical examples in two-hadron systems is $\Lambda(1405)$
as a quasi-bound state of $\bar KN$~\cite{Dalitz:1959dn}. For mesonic 
resonances, the scalar mesons, $f_{0}(980)$ and $a_{0}(980)$, are also
the candidates of the hadronic molecular states~\cite{Weinstein:1982gc}.
Baryon resonances as three-hadron systems have been also investigated
theoretically for systems of $\pi K N$~\cite{Bicudo:2003rw,LlanesEstrada:2003us,Kishimoto:2003xy}, $\pi \bar K N$~\cite{MartinezTorres:2007sr} and 
$\bar K \bar K N$~\cite{ikeda07-jps,KanadaEn'yo:2008wm}.
Based on the idea to regard $\Lambda(1405)$ as a $\bar KN$ quasi-bound 
state~\cite{akaishi02,yamazaki02},
bound systems of a few nucleons with anti-kaon were investigated in 
Refs.~\cite{akaishi02,yamazaki02,Yamazaki:2003hs,Yamazaki:2007hj,shevchenko07,ikeda07,yamazaki07,dote08}.

Recently a baryonic resonance with $J^P=1/2^+$
and $S=-2$ composed by $\bar K \bar K N$ has been studied in details  
by the authors in Ref.~\cite{KanadaEn'yo:2008wm} based on 
three-body calculation with attractive $\bar K N$ interactions 
given by Refs.~\cite{akaishi02,yamazaki07,hyodo07}. 
In this system, the anti-kaons play unique roles, because they
have enough attraction with the nucleon to form a quasi-bound state as 
$\Lambda(1405)$ and possess so heavy mass to provide small
kinetic energy in the $\bar K \bar K N$ system. 
The quasi-bound state of $\bar K \bar K N$ has a characteristic structure 
that one of the anti-kaons forms $\Lambda(1405)$ with the nucleon
($\Lambda(1405)$-cluster) as seen also in $K^{-} pp$
system~\cite{Yamazaki:2007hj}, and  
the other anti-kaon  spreads for long distance.
This structure is caused by strong $\bar K N$ attraction with $I=0$.

In this paper,  
we explore quasi-bound states of the $K \bar K N$ system with $I=1/2$ and 
$J^{P}=1/2^{+}$, assuming that the $\bar K N$ and $K\bar K$ systems
have enough attractions to form quasi-bound states of $\Lambda(1405)$ 
in $I=0$ and $f_{0}(980)$ ($a_{0}(980)$) in $I=0$ ($I=1$), respectively.
We use
the effective interactions of $\bar K N$ extracted by
Akaishi-Yamazaki~(AY)~\cite{akaishi02,yamazaki07} and Hyodo-Weise~(HW)~\cite{hyodo07} in phenomenological way and chiral dynamics, respectively.
These interaction provide 
$\Lambda(1405)$ as a quasi-bound state with $I=0$ and also weak
attraction in the $I=1$ channel. 
The effective $K\bar{K}$ interactions are adjusted to reproduce 
the masses and the widths of $f_0(980)$ and $a_0(980)$ as the $K\bar{K}$
molecular states.
The $KN$ interactions are known to have strong repulsion in $I=1$ channel.
We use the $KN$ potential fitted by observed scattering lengths.

The ``fate" of the $K\bar KN$ molecular state strongly depends on its binding 
energy.  If the energy of the $K\bar KN$ system is above the lowest threshold 
of the subcomponents, the $K\bar KN$ states can decay to the subcomponents
and the width gets very large. 
If the $K\bar KN$ state is bound with moderate binding energy 
below all the thresholds of $\Lambda(1405)$+$K$,
$f_{0}(980)$+$N$ and $a_{0}(980)$+$N$, the state is quasi-stable 
against these decay modes and has comparable decay width with those of 
the two-particle subsystems. For deeply bound $K\bar KN$ system, 
since the constituents largely overlap each other, the molecular picture may be
broken down and two-body decays are enhanced.

Having strong attractions in $\bar KN$ and $K\bar K$ subsystems,
it is naturally expected that $K\bar K N$ forms a hadron molecule below the 
thresholds of $\Lambda(1405)$-$K$ and $f_{0}$($a_{0})$-$N$.
The question arising here is 
whether or not the attractions 
are so strong that the hadronic molecular picture breaks down in 
deeply bound state and the quasi-bound state has large width, 
or in opposite direction,
whether or not the repulsion of $KN$ is too strong
for spoiling the bound state.

In Sec.~\ref{sec:formulation}, we describe the framework 
of the present calculations. We apply a variational approach with 
a Gaussian expansion method~\cite{Hiyama03} to 
solve the Schr\"odinger equation of the three-body system.
By treating the imaginary potentials perturbatively,  
we find the $K\bar{K}N$ quasi-bound state. 
In Sec.~\ref{sec:results}, we present our results of the three-body
calculation. In analysis of the wave functions,
we discuss the structure of the $K\bar{K}N$ state.
Section~\ref{sec:summary} is devoted to summary of this work. 

\section{Formulation} \label{sec:formulation}

We apply a non-relativistic three-body potential model for 
the $K\bar{K}N$ system. 
The effective two-body interactions are given
in local potential forms.
The $K\bar{K}N$ wave function 
is calculated by solving  Schr\"odinger equation
with a Gaussian expansion method for the three-body system.
In this section, we briefly explain the formulation  and 
interactions used in the present work.  The details of the formulation 
and the $\bar K N$ interaction are discussed in 
Ref.~\cite{KanadaEn'yo:2008wm}.

\subsection{Hamiltonian}

\begin{figure}[t]
\centerline{\includegraphics[width=7.5cm]{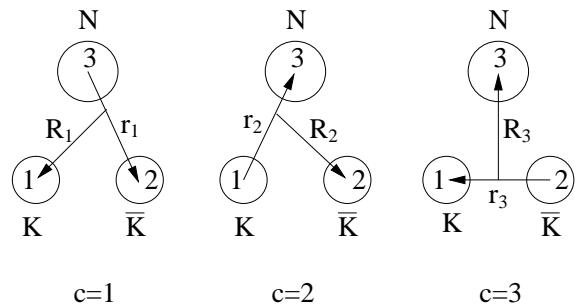}}
\caption{\label{fig:jacobi}
Three Jacobian coordinates of the $K\bar{K}N$ system.
}
\end{figure}

In the present work, the Hamiltonian for 
the $K\bar{K}N$ system is given by
\begin{equation}\label{eq:hamiltonian}
H=T+V_{\bar{K}N}(r_1)+V_{KN}(r_2)+V_{K\bar{K}}(r_3), 
\end{equation} 
with the kinetic energy $T$, the effective $\bar{K}N$ interaction
$V_{\bar{K}N}$, the $KN$ interaction $V_{KN}$ and the 
$K\bar{K}$ interaction $V_{K\bar{K}}$.
These interactions are given by $\ell$-independent  local potentials as functions of 
$\bar{K}$-$N$, $K$-$N$ and $K$-$\bar K$ distances, 
$r_{1}$, $r_{2}$ and $r_{3}$ defined by  
$r_{1}=|{\bf x}_{\bar K}-{\bf x}_N|$, $r_{2}=|{\bf x}_N-{\bf x}_K|$ and   
$r_{3}=|{\bf x}_K-{\bf x}_{\bar K}|$, respectively,  
with  spatial coordinates ${\bf x}_{K}$, ${\bf x}_{\bar K}$, ${\bf x}_N$ for
the kaon,  the anti-kaon and the nucleon. For convenience,
we introduce Jacobian coordinates,
${\bf r}_c$ and ${\bf R}_c$, in three rearrangement 
channels $c=1,2,3$ as shown in Fig.~\ref{fig:jacobi}.
We assume isospin symmetry in the effective interactions, and we also
neglect the mass differences among $K^{\pm}$, $\bar K^{0}$ and $K^0$, 
and that 
between proton and neutron by
using the averaged masses, $M_K=495.7$ MeV and $M_N=938.9$ MeV. 
We do not consider three-body forces nor transitions to two-hadron decays,
which will be important if the constituent hadrons are localized in a small region. 

The kinetic energy $T$ is simply given by the Jacobian coordinates 
with one of the rearrangement channels as 
\begin{equation}
T\equiv \frac{-1}{2\mu_{r_c}}\nabla_{r_c}^2
+\frac{-1}{2\mu_{R_c}}\nabla_{R_c}^2,
\end{equation} 
with the reduced masses $\mu_{r_c}$ and $\mu_{R_c}$ 
for the corresponding configuration,
for instance, 
$\mu_{r_{1}}=M_K M_N/(M_K+M_N)$ and 
$\mu_{R_{1}}=M_K (M_K+M_N)/(2M_K+M_N)$ for the rearrangement channel 
$c=1$.

The effective interactions, $V_{\bar KN}$, $V_{KN}$ 
and $V_{\bar K \bar K}$, are obtained by 
$s$-wave two-body scattering with isospin symmetry.
The explicit expression of the effective interactions  will be given 
in Sec.~\ref{sec:effint}.
Open channels of $\bar K N$ and $K\bar K$ ($\pi \Lambda$ and $\pi \Sigma$ for $\bar K N$, and
$\pi \pi$ and $\pi\eta$ for $K\bar K$) are implemented effectively to the imaginary 
parts of the interactions $V_{\bar KN}$ and $V_{K\bar K}$. 
Consequently, the Hamiltonian~(\ref{eq:hamiltonian}) is not 
hermitian. 
In solving Schr\"odinger equation for $K\bar{K}N$,
we first take only the real part of the potentials 
and obtain wavefunctions in a variational approach.
Then we calculate bound state energies $E$ as expectation values 
of the total Hamiltonian~(\ref{eq:hamiltonian}) with 
respect to the obtained wave functions.
The widths of the bound states are evaluated by the imaginary part of 
the complex energies as $\Gamma=-2\, {\rm Im}E$.

\subsection{Effective interactions}
\label{sec:effint}
In this subsection, we explain the details of the effective interactions of the $\bar KN$, $K \bar K$ and $KN$ two-body subsystems in our formulation.  
The interaction parameters and the properties of the two-body 
subsystems are listed in table \ref{tab:interactions}.

\subsubsection{$\bar KN$ interaction} 
In this work, 
we use the same $\bar KN$ potential as used in 
Ref.\cite{KanadaEn'yo:2008wm}
for the $\bar{K}\bar{K}N$ calculations.
We consider two different effective  $\bar KN$ 
interactions to estimate theoretical uncertainties.
These two interactions were derived in different ways.
One of the $\bar K N$ interaction that we use is given by Hyodo and Weise in Ref.~\cite{hyodo07}
and was derived  based on the chiral unitary approach for $s$-wave scattering amplitude with strangeness $S=-1$.
The other interaction is Akaishi-Yamazaki (AY) potential derived phenomenologically
by using $\bar KN$ scattering and kaonic hydrogen data and reproducing the
$\Lambda(1405)$ resonance as a $K^-p$ bound state at 1405 MeV~\cite{akaishi02,yamazaki07}.
Both $\bar KN$ interactions have so strong 
attraction in $I=0$ as to provide the $\Lambda(1405)$
as a quasi-bound state of the $\bar KN$ system,
and have weak attraction in $I=1$. 
Hereafter we refer the quasi-bound $\bar KN$ state as \KNsing.

The potential is written in a one-range Gaussian form as
\begin{eqnarray}
V_{\bar{K}N} &=& U_{\bar{K}N}^{I=0}\exp\left[ -(r/b)^2\right]P_{\bar{K}N}(I=0)
\nonumber \\ && 
+U_{\bar{K}N}^{I=1}\exp\left[ -(r/b)^2\right]P_{\bar{K}N}(I=1),
\end{eqnarray}
with the isospin projection operator $P_{\bar{K}N}(I=0,1)$ and
the range parameter $b$. The potential depth $U_{\bar KN}^{I}$ 
are given in a complex number reflecting the effects of the open 
channel of $\pi \Sigma$ and $\pi \Lambda$. The numbers for $b$ and 
$U_{\bar KN}^{I}$  are given in Table~\ref{tab:interactions}.
For the Hyodo-Weise potential, we use the parameter set referred 
as HNJH in Ref.~\cite{hyodo07}, which was obtained by
the chiral unitary model with the parameters of Ref.~\cite{Hyodo03}.
We refer this potential as ``HW-HNJH potential''.
The energy of the HW-HNJH potential is fixed at $\omega = M_{K} + M_{N}-11$
MeV which is the resonance position of $\Lambda(1405)$, since the 
energy dependence in the potential is small in the region of interest and 
it was found in Ref.~\cite{KanadaEn'yo:2008wm}
that the results were not sensitive to the choice of the energy
for the $\bar K \bar K N$ system.

The important difference between the two interactions 
is the binding energy of the $\bar K N$ system. 
In chiral unitary approaches for the meson-baryon interactions, 
the $\Lambda(1405)$ resonance is described 
as a $\bar{K}N$ quasi-bound state~\cite{Hyodo:2008xr} 
located at $\omega\sim 1420$ MeV in $\bar KN$ scattering 
amplitude~\cite{Jido:2003cb}. This is a consequence of 
the double pole nature that $\Lambda(1405)$ is 
described by superposition of two poles as found in 
Refs.~\cite{Oller:2000fj,Jido:2002yz,Jido:2003cb}.
For the AY potential, the $\Lambda(1405)$ resonance
was reproduced at $\sim$ 1405 MeV as PDG reported. 
Thus, the AY potential has stronger attraction in $I=0$ 
than the HW-HNJH potential. The properties of the 
$\bar KN$ two-body system obtained by these potentials
are summarized in Table~\ref{tab:interactions}.

\subsubsection{$K\bar K$ interaction} \label{sec:KKint}
The $K \bar K$ interaction is derived in the present work
under the assumption that $K\bar K$ forms quasi-bound
states in $I=0$ and $I=1$, which correspond to 
$f_0(980)$ and $a_0(980)$, respectively. 
Thus, 
we use strong effective single-channel $K\bar K$ interactions 
which reproduce the masses and widths of 
$f_0(980)$ and $a_0(980)$ as the quasi-bound 
$K\bar K$ states. 
We refer the quasi-bound 
$K\bar K$ states as \KK{0,1}.

We take the one-range Gaussian form, 
\begin{equation}
V^{I=0,1}_{K\bar K}(r)=U^{I=0,1}_{K\bar K}
\exp\left[ -(r/b)^2\right]P_{K\bar K}(I=0,1),
\end{equation}
where the range parameter $b$ is chosen to be the same value 
as that of the $\bar{K}N$ interaction.
We adjust the strength $U^{I=0,1}_{K\bar K}$ to fit the 
$f_0(980)$ and $a_0(980)$ 
masses and the widths with the energies of two-body calculations
of the $K\bar K$ system. The particle data group (PDG) reports~\cite{PDG}
the $f_0(980)$ and $a_0(980)$ have 
$980\pm 10$ MeV and $984.6\pm 1.2$ MeV masses with 
the $40-100$ MeV and $50-100$ MeV widths, respectively,
in average of the compilation of the experimental data.
The dominant decay modes are $\pi\pi$ for $f_0(980)$ and 
$\pi\eta$ for $a_0(980)$.
We take the mass 980 MeV and the width 60 MeV as the inputs to
determine the $K\bar K$ interactions in 
both the $I=0$ and $I=1$ channels. Then we get 
$U^{I=0,1}_{K\bar K}=-1155-283i$ MeV for $b=0.47$ fm and
$U^{I=0,1}_{K\bar K}=-630-210i$ MeV for $b=0.66$ fm, 
by fitting the energy of the $K\bar K$ bound state to the 
meson mass and the width. We refer the former potential as ``KK(A)'' and 
the latter as ``KK(B)''.
In this phenomenological single-channel 
interaction, the effect of the two-meson decays such 
as $\pi\pi$ and $\pi\eta$ decays is 
incorporated in the imaginary part of the effective
$K\bar K$ interaction.

In the present parametrization of the $K\bar K$ potential,
we have fitted the potential strengths to reproduce the PDG values
of the $f_0(980)$ and $a_0(980)$ masses as 
bound state energies of $K\bar K$  calculated 
with the perturbative treatment of the imaginary potential.
When we directly calculate the pole position of the $K\bar K$ scattering
amplitude in Lippmann-Schwinger equation with the present potential, 
we get the value $998-32i$ MeV for the parameter set (A).
This is obtained above the $K\bar K$ threshold in the first 
Riemann sheet as a virtual state.  
This pole is consistent with  the pole position of scattering amplitude 
obtained by the chiral unitary approach~\cite{Oller:1997ti,Oller:1997tiE}.
In the chiral unitary approach, 
$s$-wave scattering amplitudes with $I=0$ and $I=1$ were
reproduced well by coupled channels of $\pi\pi$, $\pi\eta$ and $K\bar K$.  
The $f_0(980)$ and $a_0(980)$ meson are obtained as 
the resonance poles at 993.5 MeV  for $f_0(980)$ and 1009.2 MeV for $a_0(980)$~\cite{Oller:1997tiE}.
The $f_{0}$ is described dominantly by $K\bar K$ scattering, while for $a_{0}$
 the $\pi \eta$ scattering is also important as well as $K\bar K$ scattering.
 These values are slightly higher than the masses 
reported by PDG, which are 
given by the peak position of the spectra.
We will discuss ambiguity of the $K\bar K$ interactions in later section.

\subsubsection{$KN$ interaction} \label{sec:KNint}
We construct the $KN$ interaction based on observed 
$KN$ scattering lengths~\cite{KNint}. 
The $KN$ interactions are known to be strong repulsion in the $I=1$ channel
and very weak in the $I=0$ channel.
The experimental values of the 
scattering lengths for the $I=0$ and $I=1$ channels are
$a^{I=0}_{KN}=-0.035$ fm and $a^{I=1}_{KN}=-0.310\pm 0.003$~fm~\cite{KNint}. 
In the present calculation, we assume no interaction in the $I=0$ channel.
For the $I=1$ channel, we use phenomenological 
interaction with the one-range Gaussian form again,
\begin{equation}
V^{I=1}_{KN}(r)=U^{I=1}_{KN}
\exp\left[ -(r/b)^2\right]P_{NK}(I=1),
\end{equation}
where the range parameter $b$ is chosen to be the same value 
as that of the $\bar{K}N$ interaction. 
We adjust the strength $U^{I=1}_{KN}$ to reproduce the experimental 
scattering strength and obtain $U^{I=1}_{KN}=820$ MeV and 
$U^{I=1}_{KN}=231$ MeV for $b=0.47$ fm and $b=0.66$ fm,
respectively. We refer the former parametrization as ``KN(A)''
and the latter one as ``KN(B)''.

\begin{table}[ht]
\caption{\protect\label{tab:interactions} 
The interaction parameters and the properties of two-body systems.
The energies ($E$) are evaluated from the corresponding two-body
threshold. They are calculated by treating the imaginary 
part of the two-body potentials perturbatively. We also list 
the root-mean-square two-body
distances of the $\bar K N(I=0)$, $K\bar K(I=0)$ and $K\bar K(I=1)$ states,
which correspond to $\Lambda(1405)$ and $f_0(980),a_0(980)$, respectively.
For the $K\bar K$ interactions, we show the scattering lengths obtained 
in the present parameters. 
}
\begin{tabular}{ccc}
\hline
 &\multicolumn{2}{c}{parameter set of interactions}\\
	&	(A)	&  (B)	\\
 & &  \\
\hline
$\bar K N$	&	HW-HNJH	&	AY	\\
 & &  \\
$b$ (fm)	&0.47 & 0.66 	\\
$U^{I=0}_{KN}$ (MeV)	&	$-908-181i$	&	$-595-83i$	\\
$U^{I=1}_{KN}$ (MeV)	&	$-415-170i$	&	$-175-105i$	\\
 & & \\
$\bar K N(I=0)$ state	&		&		\\
Re$E$ (MeV)	&	-11 	&	-31 	\\
Im$E$ (MeV)	&	-22 		&	-20 	\\
$\bar K$-$N$ distance  (fm) 	&	1.9 	&	1.4 	\\
 & & \\
\hline
$K\bar K$	& KK(A)	&	KK(B)	\\
 & &  \\
$b$ (fm)	&	0.47 		&	0.66 	\\
$U^{I=0,1}_{K\bar K}$ (MeV)	&	$-1155-283i$		&	$-630-210i$	\\
 & & \\
$K\bar K (I=0,1)$ state	&	&		\\
Re$E$ (MeV) 	&	-11 	&	-11 	\\
Im$E$ (MeV) 	&	-30 	&	-30 	\\
 $K$-$\bar K$ distance (fm) 	& 	2.1 		&	2.2 	\\
 & &  \\
\hline
$KN$	& KN(A)		&	KN(B)	\\
 & & \\
$b$ (fm)	&   0.47 	&	0.66 	\\
$U^{I=0}_{KN}$ (MeV)	& 0 		&	0 	\\
$U^{I=1}_{KN}$ (MeV)	& 820 	&	231 	\\
 & & \\
$a^{I=0}_{KN}$ (fm)	& 	0 		&	0 	\\
$a^{I=1}_{KN}$	(fm) &$-$0.31 	&	$-$0.31 \\
 & & \\
\hline
\end{tabular}
\end{table}

\subsection{Three-body wave function}\label{subsec:wf}

The three-body $K\bar{K}N$ wave function $\Psi$
is described 
as a linear combination of amplitudes 
$\Phi^{(c)}_{I_{K\bar K}}({\bf r}_{c}, {\bf R}_{c})$
of three rearrangement channels 
$c=1,2,3$ (Fig.~\ref{fig:jacobi}).
In the present calculation, we take the model space limited to
$l_c=0$ and $L_c=0$ of the orbital-angular momenta for the
Jacobian coordinates ${\bf r}_c$ and ${\bf R}_c$ in the 
channel $c$ owing to the fact that 
the effective local potentials used in the
present calculations are derived in consideration of the $s$-wave
two-body dynamics. 
Then the  wave function of the $K\bar{K}N$ system 
with $I=1/2$ and $J^P=1/2^{+}$ is written as
\begin{equation}
\Psi=\sum_{c,I_{K\bar K}} 
\Phi^{(c)}_{I_{K\bar K}}({\bf r}_c, {\bf R}_c) 
\left[[K\bar{K}]_{I_{K\bar K}}N\right]_{I=1/2} 
\label{eq:wavefunc}
\end{equation}
where the $\left[[K\bar{K}]_{I_{K\bar K}}N\right]_{I=1/2}$ specifies 
the isospin configuration of the wave function 
$\Phi^{(c)}_{I_{K\bar K}}({\bf r}_c, {\bf R}_c)$, 
meaning that the total isospin $1/2$ for the $K \bar K N$ system is 
given by combination of total isospin $I_{K\bar K}$ for the $K \bar K$ 
subsystem and isospin $1/2$ for the nucleon.

The wave function of the $K\bar{K}N$ system is obtained by solving  
the Schr\"odinger equation,
\begin{equation}
\left[ T+V_{\bar{K}N}(r_1)+V_{KN}(r_2)
+V_{K\bar{K}}(r_3)-E\right]\Psi=0.
\end{equation}
In solving the Schr\"odinger equation for 
the $K\bar{K}N$ system,
we adopt the Gaussian expansion method for three-body systems given in
Ref.~\cite{Hiyama03} as same way as done in 
Ref.\cite{KanadaEn'yo:2008wm} with the parameters  $r_{\rm min},R_{\rm min}=0.2$ fm and $r_{\rm max},R_{\rm max}=20$ fm,
and $n_{\rm max},N_{\rm max}=15$ for all the channels, $c=1,2,3$.  
We treat the imaginary part of the potentials perturbatively.
We first calculate the wave function for the real part of the 
Hamiltonian ($H^{\rm Re}$)
with variational principle
in the model space of the Gaussian expansion. 
After this variational calculation, we take 
the lowest-energy solution.
The binding energy $B(K\bar{K}N)$ of the three-body
system is given as $B(K\bar{K}N)=-E^{\rm Re}$.

Next we estimate the imaginary part of the energy 
for the total Hamiltonian 
by calculating the expectation value of the imaginary part of the
Hamiltonian with the obtained 
wave function $\Psi$:
\begin{equation}
E^{\rm Im}=\langle \Psi| {\rm Im}V_{\bar{K}N}+{\rm Im}V_{K\bar K}|\Psi \rangle.
\end{equation}
The total energy is given as $E=E^{\rm Re}+E^{\rm Im}i$, and the 
decay width is estimated as 
$\Gamma=-2 E^{\rm Im}$. In the present calculation, 
we have only three-body decays such as
$\pi \Sigma \bar K$, $\pi \Lambda \bar K$, $\pi \pi N$
and $\eta \pi N$ decays for the $K\bar KN$ state
by the model setting. 

The perturbative treatment performed above is justified qualitatively
in the case of $|\langle \Psi| {\rm Im}V|\Psi \rangle| \ll
|\langle \Psi| {\rm Re}V|\Psi \rangle|$. 
In the two-body systems, $\bar{K}N$ and $K\bar K$, we find that 
this condition is 
satisfied reasonably, observing  that 
$|\langle {\rm Im}V_{\bar{K}N} \rangle|\sim 20$ MeV is much smaller than
$|\langle {\rm Re}V_{\bar{K}N} \rangle|\sim 100$ MeV,
and $|\langle {\rm Im}V_{K\bar{K}} \rangle|\sim 30$ MeV is also much 
smaller than $|\langle {\rm Re}V_{K\bar{K}} \rangle|\sim 100$ MeV.
Also in the case of the $K\bar{K}N$ system, 
it is found that the absolute values of the
perturbative energy 
$|\langle \Psi| {\rm Im}V |\Psi \rangle|
\sim 50 $ MeV 
is much smaller than the real potential energy
$|\langle \Psi| {\rm Re}V|\Psi \rangle| \sim 200$ MeV
in the present calculations.

We also calculate quantities characterizing the structure of the three-body system,
such as spatial configurations of the constituent particles and 
probabilities to have specific isospin configurations. 
These values are calculated as expectation values of the wave functions.

The root-mean-square (r.m.s.)~radius of the $K\bar K N$ state
is defined as the average of the distribution of $K$, $\bar K$ and $N$ by
\begin{equation}
r_{K\bar KN}=
\sqrt{\left \langle \Psi\left|
{\textstyle \frac{1}{3}}({\bf x}^2_{K}+{\bf x}^2_{\bar K}+{\bf x}^2_N)\right|
\Psi \right\rangle}, \label{eq:radKKN}
\end{equation}
which is measured from the center of mass of the three-body system.
We also
calculate the r.m.s.~values of the relative distances between two particles, 
\begin{eqnarray}
d_{\bar K N} &=& \sqrt{\left \langle \Psi\left|
{\bf r}_{1}^2\right|
\Psi \right\rangle},   \label{eq:distKbN} \\
d_{K N} &=& \sqrt{\left \langle \Psi\left|
{\bf r}_{2}^2\right|
\Psi \right\rangle}, \label{eq:distKN} \\
d_{K \bar K} &=& \sqrt{\left \langle \Psi\left|
{\bf r}_{3}^2\right|
\Psi \right\rangle}. \label{eq:distKKb}
\end{eqnarray}
Here $r_1$, $r_2$ and $r_3$ are the $\bar KN$ distance, 
$KN$ distance and the $K\bar K$ distance, respectively.

We also introduce the probabilities for the three-body system 
to have the isospin $I_{K\bar K}$ states as
\begin{equation}
\Pi \left(\left[K \bar K\right]_{I_{K \bar K}}\right)
\equiv \left\langle \Psi \left|P_{\bar K}(I_{K \bar K})\right|
\Psi \right\rangle, \label{eq:probKK}
\end{equation}
where $P_{K\bar K}(I_{K\bar K})$ is the projection operator
for the isospin configuration
$\left[[K\bar{K}]_{I_{K\bar K}}N\right]_{I=1/2}$, as introduced before.
We calculate the probabilities that the three-body system has
the isospin configurations of 
$\left[[\bar{K}N]_{I_{\bar KN}}\right]_{I=1/2}$, where
the total isospin $1/2$ 
is given by combination of total isospin $I_{\bar KN}$ for the $\bar K N$
subsystem and the kaon isospin $1/2$:
\begin{equation}
\Pi \left(\left[\bar KN\right]_{I_{\bar KN}}\right)
\equiv \left\langle \Psi \left|P_{\bar K N}(I_{\bar KN})\right|
\Psi \right\rangle,
\end{equation}
where $P_{\bar K N}(I_{\bar KN})$ is again the isospin projection operator.

\section{Results} \label{sec:results}

In this section, we show the results of 
investigation of the $K \bar K N$ system with $I=1/2$ and 
$J^P=1/2^+$. We consider two parameter sets (A) and (B) for the
two-body interactions listed in Table~\ref{tab:interactions}. 
For the $\bar K N$ interactions, 
we use (A) the HW-HNJH potential 
and (B) the AY potential.
For the $K\bar K$ and $KN$ interactions, we use the 
phenomenological interactions derived in Sec.~\ref{sec:KKint} 
and~\ref{sec:KNint}: KK(A) and KN(A) for set (A),
and KK(B) and KN(B) for set (B). 
In addition, we study the effect of the $KN$ repulsion 
by switching off the $KN$ interaction in the parameter sets
(A) and (B).

\subsubsection{Properties of $K \bar KN$ state}

\begin{figure}[th]
\centerline{\includegraphics[width=7.5cm]{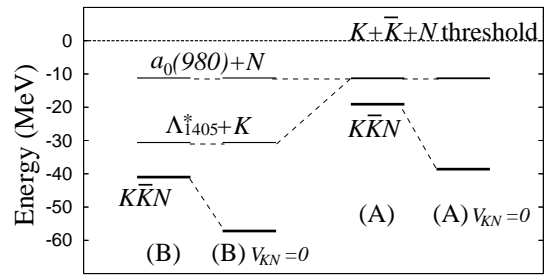}}
\caption{\label{fig:spe} Level structure of the $K \bar{K}N$ system
calculated with (a)~the HW-HNJH potential
and (b)~the AY potential. The energies are measured from 
the $K$+$\bar K$+$N$ threshold located at 1930 MeV.
The $K \bar K N$ bound state is denoted by $K \bar K N$.
The calculated  thresholds of the two-body decays to 
\KK{0,1}$+N$ and \KNsing$+K$ are denoted by
$a_{0}(980)+N$ and $\Lambda^{*}_{1405}+K$, respectively. 
The results obtained without the $KN$ repulsion are also shown. }
\end{figure}

First of all, we find that, 
in both calculations (A) with the HW-HNJH and 
(B) with the AY potentials,
the $K\bar{K}N$ bound 
state is obtained below all threshold energies of 
the \KNsing$+K$, \KK{0}$+N$ and \KK{1}$+N$ channels, which correspond to 
the $\Lambda(1405)+K$, $f_0(980)+N$ and $a_0(980)+N$ states, 
respectively.\footnote{In the present calculation, because the $K\bar K$ 
interaction is adjusted to 
reproduce the $f_{0}$ and $a_{0}$ scalar mesons having 
the same mass and width, it is independent of the total isospin of the $K\bar K$
subsystem and the thresholds of $f_0(980)+N$ and $a_0(980)+N$
are obtained as the same value. }
This means that the obtained bound state is stable against breaking up to
the subsystems. 
We show 
the level structure of the 
$K \bar K N$ system measured from the $K+\bar K+N$ threshold
in Fig.~\ref{fig:spe}. 
The values of the real and imaginary parts of the obtained energies are
given in~Table \ref{tab:energy}. The imaginary part of the energy 
is equivalent to the half width of the quasi-bound state. 
The contribution of each decay mode to the imaginary energy is shown 
as an expectation value of the imaginary potential  $\langle {\rm Im} V \rangle$ 
and the results obtained without the $KN$ interaction are also given.
For the  HW-HNJH potential, since the original HW-HNJH potential 
is moderately dependent on the energy of the $\bar KN$ system, 
we have calculated the bound state energy with the $\bar KN$ potential
for the $\bar KN$ energy at $\omega = M_{K}+M_{N}$ and 
found that  the energy ($\omega$) 
dependence of the HW-HNJH potential is small in the result.

\begin{table}[th]
\caption{Energies of the $K\bar KN$ states calculated with the 
parameter sets (A) and (B) given in Table~\ref{tab:interactions}. 
The results without the $KN$ repulsive interaction are also shown. 
Contributions of $V^{I=0,1}_{\bar KN}$ and
$V^{I=0,1}_{K\bar K}$ to the imaginary energy are separately listed.
\protect\label{tab:energy} }
\begin{tabular}{lcccc}
\hline
parameter set	&	(A)	&	(A)	&	(B)	&	(B)	\\
$V_{\bar KN}$	&	HW-HNJH	&	HW-HNJH	&	AY	&	AY	\\
$V_{KN}$ 	&	on	&	off	& on	&	off	\\
\hline
Re$E$ (MeV)	&$	-19 	$&$	-39 	$&$	-41 	$&$	-57 	$\\
Im$E$ (MeV)	&$	-44 	$&$	-72  	$&$	-49 	$&$	-63 	$\\
$\langle {\rm Im}V^{I=0}_{\bar KN} \rangle$ (MeV)	&$	-17 	$&$	-30 	$&$	-19 	$&$	-23 	$\\
$\langle {\rm Im}V^{I=1}_{\bar KN} \rangle$ (MeV)	&$	-1 	$&$	0  	$&$	0 	$&$	0 	$\\
$\langle {\rm Im}V^{I=0}_{K\bar K} \rangle$ (MeV)	&$	-1 	$&$	-10 	$&$	-4 	$&$	-10 	$\\
$\langle {\rm Im}V^{I=1}_{K\bar K} \rangle$ (MeV)	&$	-25 	$&$	-31 	$&$	-25 	$&$	-31 	$\\
\hline
\end{tabular}
\end{table}

Let us discuss first the results with the $KN$ interaction in detail. 
The binding energy of the  $K \bar K N$ state measured 
from the three-body $K+\bar K+N$ threshold is larger in the result with (B) 
than that with (A), as found to be $-19$ MeV and $-41$ MeV in the cases of 
(A) and (B), respectively. 
This is because
the AY potential gives a deeper binding of the \KNsing\ state
than the HW-HNJH potential due to the stronger $\bar KN$ attraction. 
These values have meaning just for the position of the quasi-bound state
in spectrum. It is more physically important that
the $K \bar K N$ bound state appears about 10 MeV below
the lowest two-body threshold, \KNsing$+K$, 
in both cases (A) and (B). 
This energy is compatible to nuclear many-body system, 
and it is considered to be weak binding energy
in the energy scale of hadron system.
This weak binding system has the following significant feature.
The width of the $K \bar K N$ state is estimated to be 
$\Gamma \sim 90$ MeV from the imaginary part of the energy.
Comparing the results of the $K \bar K N$ with 
the properties of the two-body subsystems shown in 
Table~\ref{tab:interactions}, it is found that the real and imaginary 
energy of the $K \bar K N$ state is almost given  by the sum of 
those of $\Lambda(1405)$ and 
$a_0(980)$ (or $f_0(980)$), respectively. 
This indicates that two subsystems, $\bar K N$ and $K\bar K$,
are as loosely bound in the three-body system as they are in two-body system.

The decay properties of the $K \bar K N$ state can be discussed by
the components of the imaginary energy.
As shown in Table~\ref{tab:energy}, 
among the total width $\Gamma=-2E^{\rm Im}\sim 90$ MeV, 
the imaginary potentials of the $\bar KN$ with $I=0$ 
and the $K\bar K$ with $I=1$ give large contributions
as about $40$ MeV and $50$ MeV, respectively.
The former corresponds to the $\Lambda(1405)$ decay channel
and gives  the $\bar KN\rightarrow\pi\Sigma$ decay mode with $I=0$. 
The latter is given by the $a_0(980)$ decay, which is dominated 
by $K\bar K \rightarrow \pi \eta$. 
Contrary, the $\bar KN$ $(I=1)$ 
and the $K\bar K$ $(I=0)$ interactions provide only small
contributions to the imaginary energy. This is because, as 
we will see later, the $\bar K N$ subsystem is dominated by the
$I=0$ component due to the strong $\bar KN$ attraction and 
the $K \bar K$ subsystem largely consists of  the $I=1$ component
as a result of the three-body dynamics. 
The small contributions of the $\bar KN$ $(I=1)$ 
and the $K\bar K$ $(I=0)$ interactions to the imaginary energy
implies that the decays to $\pi\Lambda K$ and $\pi\pi N$ are 
suppressed. 
Therefore, we conclude that the
dominant decay modes of the $K \bar K N$ state are 
$\pi\Sigma K$ and $\pi\eta N$.   This is one of the important 
characters of the $K \bar K N$ bound system. 

Although the obtained $K \bar K N$ state is located below
the thresholds of  $\Lambda(1405)+K$, 
$f_0(980)+N$ and $a_0(980)+N$, 
there could be a chance to access the $K \bar K N$ state energetically  
by observing the $\Lambda(1405)+K$, $f_0(980)+N$ and $a_0(980)+N$ 
channels in the final states,  because these resonances have as large widths as  
the  $K\bar K N$ state. 
Since, as we will show later,  the $K \bar K N$ state
has the large $\Lambda(1405)+K$ component, 
the $K \bar K N$ state 
could be confirmed in its decay to  $\Lambda(1405)+K$
by taking coincidence of the $\Lambda(1405)$
out of the invariant mass of $\pi\Sigma$ and the three-body 
invariant mass of the $\pi\Sigma K$ decay.

Here we comment on theoretical uncertainty of the energy 
of the  $K \bar K N$ state. 
In the present calculations, the $K\bar K$ interactions are 
obtained under the assumption that the $K \bar K$ attractive 
potentials provide $f_0(980)$ and $a_0(980)$ as quasi-bound 
states and are phenomenologically adjusted to reproduce 
the masses and the widths of $f_0(980)$ and $a_0(980)$.
As discussed above, in the present result, 
the $K\bar K$ interaction  with $I=1$
gives the dominant contribution to the total width of the $K \bar K N$ state. 
We estimate theoretical uncertainty of the width of 
the $K \bar KN$ state by changing the inputs of the $a_{0}(980)$ width
in the range from $\Gamma=50$ to $100$ as reported in PDG.  
We obtain the $\Gamma=80-130$ MeV for the $K \bar K N$ state. 
We also find that the $K \bar K N$ state becomes unbound if the $K\bar K$
interaction with $I=1$ is less attraction than 70\% of the present values,
in which $K\bar K$ with $I=1$ is not bound in the two-body system.

Finally we discuss the role of the $KN$ repulsion in the
$K \bar K N$ system. In Fig.~\ref{fig:spe} and 
Table \ref{tab:energy}, we show the results 
calculated without the $KN$ interaction. We find that 
the binding energy of the $K \bar KN$ state is 20 MeV larger than 
the case of the calculation with the $KN$ repulsion in both  (A) and (B) cases,
and that the absolute value of the imaginary energy also becomes larger
as $E^{\rm Im}=-72$ MeV and $-63$ MeV for (A) and (B), respectively. 
These values correspond to the $\Gamma=130-140$ MeV width for the
$K \bar K N$ state. The reason that the three-body system without the $KN$ 
interaction has the more binding and the larger
width is as follows. In general, the three-body system has less kinetic energy 
than the two-body system because of larger reduced mass in the 
three-body system.  With less kinetic energy the system can localize more.
As a result of the localization of the system, 
the system can gain more potential energy and larger imaginary energy
in the case of no $KN$ interaction than the case with the $KN$ repulsion. 
In other words, thanks to the $KN$ repulsion, 
the $K \bar K N$ state is weakly bound  and its width is suppressed to be
as small as the sum of the widths of the subsystems. 

\subsubsection{Structure of $K\bar{K}N$ state}

\begin{table}[t]
\caption{Isospin and spatial structure of the $K\bar KN$ state with the 
parameter sets (A) and (B) given in Table~\ref{tab:interactions}. 
The results without the $KN$ repulsive interaction are also shown.
The r.m.s.\ radius of the $K$, $\bar K$ and $N$ distribution, and
the r.m.s.\ values for the $\bar K$-$N$, $K$-$\bar K$ and 
$K$-$N$ distances are listed. 
The isospin components of the subsysytems, $\bar KN$ and
$K\bar K$ are also shown. The detailed definitions are described in 
Sec.~\ref{subsec:wf}.
The three-body wavefunction is obtained in  the same way as that in 
Table~\ref{tab:energy}.
\protect\label{tab:radii} }
\begin{tabular}{lcccc}
\hline
	&	(A)	&	(A)	&	(B)	&	(B)	\\
	&	HW-HNJH	&	HW-HNJH	&	AY	&	AY	\\
$V_{KN}$ 	&	on	&	off	&	on	&	off	\\
\hline
 isospin configuration & & & & \\
$\Pi \left(\left[\bar K N\right]_{0}\right)$	&	0.93 	&	1.00 	&	0.99 	&	1.00 	\\
$\Pi \left(\left[\bar K N\right]_{1}\right)$	&	0.07 	&	0.00 	&	0.01 	&	0.00 	\\
$\Pi \left(\left[K \bar K\right]_{0}\right)$	&	0.09 	&	0.25 	&	0.17 	&	0.25 	\\
$\Pi \left(\left[K \bar K\right]_{1}\right)$	&	0.91 	&	0.75 	&	0.83 	&	0.75 	\\
 spatial structure & & &  \\
$r_{K\bar KN}$ (fm) & 1.7  & 1.0  & 1.4  & 1.0  \\
$d_{\bar KN}$ (fm)	&	2.1 	&	1.3 	&	1.3 	&	1.2 	\\
$d_{K\bar K}$ (fm)	&	2.3 	&	1.4 	 &	2.1 	&	1.5 	\\
$d_{KN}$ (fm)	&	2.8 	&	1.6 	 &	2.3 	&	1.6 	\\
\hline
\end{tabular}
\end{table}

We discuss the structure of the $K\bar{K}N$ system with $I=1/2$.
For this purpose, we analyze the 
wave functions obtained in the present few-body calculation 
in terms of the spatial structure and 
the isospin configuration of the  $K\bar{K}N$
system.

We first investigate the isospin configuration of the 
$K\bar{K}N$ state. We show the isospin components of subsystems
$\bar K N$ and $K\bar K$ in Table~\ref{tab:radii}. 
It is found that the $\bar K N$ subsystem has a dominant 
$I=0$ component because of the strong $\bar K N$ interaction
in the $I=0$ channel.
In the $K\bar K$ subsystem, the $I=1$ configuration is dominant 
while the $I=0$ component gives minor contribution.
This isospin configuration is caused by the following reason. 
In both $I=0$ and $I=1$ channels, the $K\bar K$ attraction is 
strong enough to provide quasi-bound $K\bar K$ states of 
$f_0(980)$ and $a_0(980)$. In addition, 
since these scalar mesons have similar masses and
widths, the $K\bar K$ interactions in $I=0$ and $I=1$ adjusted to 
these masses and widths are similar to each other. In fact, 
we use the same parametrization for the $K\bar K$ 
interactions in the present calculation, which gives isospin-blind
potential.   
Therefore, the $\bar K N$ interaction plays a major role
to determine the isospin configuration of the $K\bar{K}N$ state.
Since the $\bar K N$ interaction has stronger attraction in the $I=0$
channel than in the $I=1$ channel, 
the system prefers to have $I=0$ in the $\bar KN$ subsystem.
If the $\bar KN$ subsystem has pure $I=0$ configuration, which is the 
case without the $KN$ repulsion, 
the $K\bar K$ subsystem should be composed by 
$I=0$ and $I=1$ with the ratio of 1:3 to have $I=1/2$ of $K\bar KN$.
Thus, the $\bar KN$ with $I=0$ 
dominates the $K\bar KN$ system, and simultaneously 
the $K\bar K$ with $I=1$ is dominant component.
The small deviation from the pure $\bar K N(I=0)$ configuration 
in the $K\bar K N$ state originates in the $KN$ repulsion.

Next we discuss the spatial structure of the $K \bar KN$ bound system. 
In Table \ref{tab:radii}, we show the root-mean-square (r.m.s.) 
radius of $K\bar K N$, $r_{K \bar K N}$ defined in Eq.~(\ref{eq:radKKN}), 
and r.m.s.~values for the $\bar K$-$N$, $K$-$\bar K$ and $K$-$N$
distances, $d_{\bar KN}$, $d_{KN}$, $d_{K\bar K}$ defined in 
Eqs.~(\ref{eq:distKbN}), (\ref{eq:distKN}) and (\ref{eq:distKKb}), respectively, 
in the $K\bar KN$ state. 
The r.m.s.~distances  of the two-body systems, 
\KNsing\ and  \KK{0,1}, are shown 
in Table~\ref{tab:interactions}.
It is interesting that the present result shows that 
the r.m.s.~$\bar K$-$N$ and $K$-$\bar K$ distances in the three-body
$K\bar KN$ state have values close to those in the quasi-bound two-body 
states, \KNsing\ and  \KK{0,1}, respectively.
This implies again that the two subsystems of the three-body state
have very similar characters with those in the isolated two-particle systems. 

The r.m.s.~$K$-$N$ distance is relatively larger 
than the r.m.s.~$\bar K$-$N$ and $K$-$\bar K$ distances due to the
$KN$ repulsion. The effect of the repulsive $KN$ interaction is important
in the present system.
Without the repulsion, we obtain smaller three-body systems as shown 
in Table~\ref{tab:radii}. Especially the distances of the two-body subsystems
are as small as about 1.5 fm, which is comparable with the sum of 
the charge radii of proton (0.8 fm) and $K^{+}$ (0.6~fm).
For such a small system, 
three-body interactions and transitions to two particles could be 
important. In addition,  
the present picture that the system is described in non-relativistic three
particles might be broken down, and one would need relativistic 
treatments and  two-body potentials with consideration of 
internal structures of the constituent hadrons.

\begin{figure}
\centerline{\includegraphics[width=4.cm]{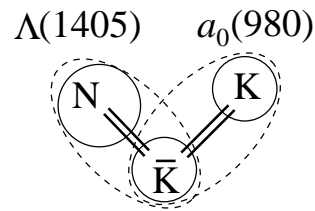}}
\label{fig:bond}
\caption{Schematic structure of the $K\bar KN$ bound system.} 
\end{figure}

Combining the discussions of the isospin and spatial structure of the
$K\bar KN$ system, we conclude that the structure of the $K \bar KN$
state can be understood  
simultaneous coexistence of $\Lambda(1405)$ and $a_{0}(980)$
clusters as shown in Fig.~\ref{fig:bond}. This does not mean
that the $K \bar KN$ system is described as superposition of 
the $\Lambda(1405)+K$ and $a_{0}(980)+N$ states, because
these states are not orthogonal to each other. The probabilities for the 
$K \bar K N$ system to have these states are 90\% as 
seen in Table~\ref{tab:radii}. It means that $\bar K$ is sheared by 
both $\Lambda(1405)$ and $a_{0}$ at the same time. 

It is interesting to compare the obtained $K \bar KN$ state with 
nuclear systems. 
As shown in Table~\ref{tab:radii}, the hadron-hadron distances in the 
$K \bar KN$ state are about 2 fm, which is as large as 
nucleon-nucleon distances in nuclei. 
In particular, in the case~(A) of the
HW-HNJH potential, the hadron-hadron distances are larger than 2 fm and 
the r.m.s.~radius of the three-body system is also  as large as 1.7 fm. This is
larger  than the r.m.s. radius 1.4 fm for $^4$He.
If we assume a uniform sphere density of the three-hadron system
with the r.m.s.~radius 1.7 fm, the mean hadron density
is evaluated as 0.07 hadrons$/$(fm$^{3}$). Thus the $K\bar KN$ state
has large spatial extent and dilute hadron density.

Order of two-body decay widths can be estimated by geometrical argument. 
Let us suppose that transitions of three-body bound state to two particles 
are induced by contact interactions. In such cases, the transition probability 
is proportional to square of density, $1/r^{6}$, 
where $r$ is the radius of the three-body system.
In the present calculation, the system of the radius is obtained as 1.7 fm
in the parameter set (A). 
Assuming a typical decay width and radius of baryon resonances as 
300 MeV and 0.8 fm, we estimate the two-body decay width as 
$300\times (0.8/1.7)^{6} \sim 3$~MeV. This is much smaller than the expected
three-body decays. 

\section{Summary} \label{sec:summary}

We have investigated the $K \bar KN$ system with $J^{p}=1/2^{+}$ 
and $I=1/2$ in non-relativistic three-body calculation.
We have used the effective $\bar K N$ potentials proposed by
Hyodo-Weise and Akaishi-Yamazaki, which reproduce the 
$\Lambda(1405)$ as a quasi-bound state of $\bar KN$. 
The $K\bar K$ interactions are determined so as
to reproduce $f_{0}(980)$ and $a_{0}(980)$ 
as quasi-bound states in $K \bar K$ with $I=0$ and
$I=1$ channels, respectively. The  potentials of  $KN$ 
 are adjusted to provide the observed $KN$ scattering lengths,
 having strong repulsion in $I=1$ and no interaction in $I=0$.
The present three-body calculation suggests that 
a weakly quasi-bound state can be formed below all threshold
energies of the $\Lambda(1405)$+$K$, $f_{0}(980)$+$N$
and $a_{0}(980)$+$N$. The calculated energies of the quasi-bound state 
are $-19$ MeV and $-41$ MeV from the $K\bar KN$ threshold
in the results with HW and AY potentials, respectively.
The width for three-hadron decays is estimated to be $90\sim 100$ MeV.
It has been found that the binding energy and the width of 
the $K\bar KN$ state 
is almost the sum of those in $\Lambda(1405)$ and $a_{0}(980)$.

Investigating the structure of the $K\bar KN$ system,
we have found that, in the $K\bar KN$ state,  
the subsystems of  $\bar KN$ and $K\bar K$ 
dominate  the  $I=0$ and $I=1$, respectively,
and that these subsystems have very similar 
properties with those in the two-particle systems. 
This leads that the $K \bar KN$ quasi-bound system
can be interpreted as coexistence state of $\Lambda(1405)$
and $a_{0}(980)$ clusters and $\bar K$ is a constituent
of both $\Lambda(1405)$
and $a_{0}(980)$ at the same time. 
As a result of this feature,  the dominant decay modes are
$\pi \Sigma K$ from the $\Lambda(1405)$ decay 
and $\pi \eta N$ from the $a_{0}(980)$ decay,
and the decays to $\pi \Lambda K$ and $\pi\pi N$ channels are 
suppressed. 

We also have found that the root-mean-square radius of the  
$K\bar KN$ state is as large as 1.7 fm and the inter-hadron 
distances are lager than 2 fm. These values are comparable to,
or even larger than, the radius of $^{4}$He and typical 
nucleon-nucleon distances in nuclei, respectively. Therefore, 
the $K\bar KN$ system more spatially extends 
compared with typical hadronic systems.  
These features are caused by weakly binding of the three hadrons,
for which the $KN$ repulsive interaction plays an important role.

\section*{Acknowledgments}

The authors would like to thank Professor~Akaishi,
Dr.~Hyodo and Dr.~Dot\'e for valuable discussions. 
They are also thankful to members of 
Yukawa Institute for Theoretical Physics (YITP)
and Department of Physics in Kyoto University, especially for
fruitful discussions.
This work is supported in part by
the Grant for Scientific Research (No.~18540263 and
No.~20028004) from Japan Society for the Promotion of Science (JSPS)
and from the Ministry of Education, Culture,
Sports, Science and Technology (MEXT) of Japan.
A part of this work is done in the Yukawa International Project for 
Quark-Hadron Sciences (YIPQS).
The computational calculations of the present work were done by
using the supercomputer at YITP.

\end{document}